# Scale-free Universal Spectrum for Atmospheric Aerosol Size Distribution for Davos, Mauna Loa and Izana

A.M.Selvam

Deputy Director (Retired)

Indian Institute of Tropical Meteorology, Pune 411006, India

Email: amselvam@gmail.com
Websites: http://amselvam.webs.com
http://www.geocities.ws/amselvam

**Abstract**

Atmospheric flows exhibit fractal fluctuations and inverse power law form for power spectra indicating an eddy continuum structure for the selfsimilar fluctuations. A general systems theory for fractal fluctuations developed by the author is based on the simple visualisation that large eddies form by space-time integration of enclosed turbulent eddies, a concept analogous to Kinetic Theory of Gases in Classical Statistical Physics. The ordered growth of atmospheric eddy continuum is in dynamical equilibrium and is associated with Maximum Entropy Production. The model predicts universal (scale-free) inverse power law form for fractal fluctuations expressed in terms of the golden mean. Atmospheric particulates are held in suspension in the fractal fluctuations of vertical wind velocity. The mass or radius (size) distribution for homogeneous suspended atmospheric particulates is expressed as a universal scale-independent function of the golden mean, the total number concentration and the mean volume radius. Model predicted spectrum is in agreement (within two standard deviations on either side of the mean) with total averaged radius size spectra for the AERONET (aerosol inversions) stations Davos and Mauna Loa for the year 2010 and Izana for the year 2009 daily averages. The general systems theory model for aerosol size distribution is scale free and is derived directly from atmospheric eddy dynamical concepts. At present empirical models such as the log normal distribution with arbitrary constants for the size distribution of atmospheric suspended particulates are used for quantitative estimation of earth-atmosphere radiation budget related to climate warming/cooling trends. The universal aerosol size spectrum will have applications in computations of radiation balance of earth-atmosphere system in climate models.

## 1. Introduction

Aerosols play an important role in the radiation budget of the earth-atmosphere system. Climate models require precise definition of aerosol size spectrum for incorporating the heat exchange processes by scattering, absorption and reflection of incident solar radiation by suspended particulates in the atmosphere. A comprehensive historical review of developments in the field of atmospheric aerosols has been given by Husar (2005). At present the statistical lognormal distribution is used as an empirical fit with arbitrary constants for single mode aerosol size distribution; multiple modes (bimodal or trimodal) are represented by superimposition of the corresponding individual lognormal distributions. The lognormal or the earlier Junge inverse powerlaw form is now established as a universal pattern for the observed aerosol size distribution at any location. The physical basis for the observed recognizable universal statistical pattern of aerosol size spectrum in the turbulent (chaotic) atmospheric flows is not yet identified. Lovejoy and Schertzer (2010) have shown conclusively that atmospheric flows exhibit selfsimilar fractal fluctuations, i.e. long-range correlations on all space-time scales exhibited as inverse powerlaw form for corresponding power spectra of meteorological parameters such as wind, temperature, pressure, etc.

Traditional meteorological theory is yet to incorporate the fractal nature of fluctuations for realistic simulation and prediction of atmospheric flows. The author (Selvam, 1990, 2007) has developed a general systems theory for the observed ordered pattern formation in turbulent atmospheric flows characterized by selfsimilar fractal fluctuations. The theory is based on classical statistical physical concepts and satisfies the principle of maximum entropy production for dynamical system in steady-state equilibrium. The model predicts the following dynamical equilibrium patterns in atmospheric flows. (i) Fractal fluctuations signify a nested continuum of vortex roll circulations with overall logarithmic spiral trajectory tracing the quasiperiodic Penrose tiling pattern for the internal structure. (ii) The probability distribution of the amplitudes as well as the variances (square of amplitude) of fractal fluctuations are represented by the same universal inverse powerlaw distribution incorporating the golden mean. Therefore fractal fluctuations signify quantumlike chaos since the property that the additive amplitude of eddies when squared represent the probability densities is exhibited by the subatomic dynamics of quantum systems such as the electron or photon. The atmospheric eddy energy (variance) spectrum is quantified by universal inverse power law distribution incorporating the golden mean (Selvam, 2009, 2011). (iii) Atmospheric particulates are held in suspension by the fractal fluctuation of wind (vertical velocity). Atmospheric aerosol mass size spectrum is expressed in terms of the universal eddy energy spectrum and is a function of the mean volume radius and total number concentration. Model predictions are in agreement with observed daily mean Aeronet aerosol size spectrum at Davos and Mauna Loa for the year 2010 and at Izana for year 2009. The paper is organized as follows. A summary of the general systems theory for atmospheric flows and aerosol size spectrum are given in Sections 2 and 3 respectively. Details of data in Section 4 is followed by results of analysis in Section 5 and conclusions in Section 6.

# 2. General Systems Theory for Fractal Fluctuations

The pioneering work by Lovejoy and his group (2010) has established conclusively that the apparently chaotic (turbulent) atmospheric flows consist of selfsimilar fractal fluctuations with eddy continuum structure exhibited in the power spectrum with inverse power law characteristics for fractal fluctuations. Inverse power law spectrum for space-time fluctuations is ubiquitous to dynamical systems in nature and signifies long-range correlations or non-local connections between the large and small-scale events in space and/or time. It is important to identify and quantify the physical laws underlying fractal fluctuations for predictability of future evolution of dynamical systems such as weather patterns. The general systems theory developed by the author is based on the simple concept (Townsend, 1956) that space-time integration of enclosed turbulent eddy fluctuations results in large eddy circulations. The above method is analogous to statistical physical procedure of obtaining bulk physical properties of say, ideal gases, by spatial integration of enclosed microphysical random molecular motions in the kinetic theory of gases The relationship between the radius $R$ and circulation speed $W$ of large eddy is related to the enclosed turbulent eddy radius $r$ and circulation speed $w_*$ is given as

$$W^2 = \frac{2}{\pi}\frac{r}{R}w_*^2 \tag{1}$$

The above equation quantifies the ordered two-stream energy flow between the larger and the enclosed smaller scales. Upward growth of large eddy occurs by latent heat energy release from microscale fractional condensation (MFC) in upward turbulent eddy fluctuations even in an unsaturated environment. From the concept of eddy growth, vigorous counter flow (mixing) characterizes the large-eddy volume. The total fractional volume dilution rate of the large eddy by vertical mixing across unit cross-section is derived from Eq. (1) and is given as follows

$$k = \frac{w_*}{\mathrm{d}W}\frac{r}{R} \tag{2}$$

In Eq. (2) $w_*$ is the increase in vertical velocity per second of the turbulent eddy due to microscale fractional condensation (MFC) process and d$W$ is the corresponding increase in vertical velocity of large eddy.

The fractional volume dilution rate $k$ is equal to 0.4 for the scale ratio ($z$) $R/r$ =10. Identifiable large eddies can exist in the atmosphere only for scale ratios more than 10 since, for smaller scale ratios the fractional volume dilution rate $k$ becomes more than half.

From Eq. (2) the following logarithmic wind profile relationship for the *ABL* is obtained

$$W = \frac{w_*}{k}\ln z \tag{3}$$

The steady state fractional upward mass flux $f$ of surface air at any height $z$ can be derived using Eq. (3) and is given by the following expression

$$f = \sqrt{\frac{2}{\pi z}}\ln z \tag{4}$$

In Eq. (4) $f$ represents the steady state fractional volume of surface air at any level z. Since atmospheric aerosols originate from surface, the vertical profile of mass and number concentration of aerosols follow the $f$ distribution.

# 3. Atmospheric aerosol size spectrum

## 3.1 Vertical variation of aerosol number concentration

The atmospheric eddies hold in suspension the aerosols and thus the size spectrum of the atmospheric aerosols is dependent on the vertical velocity spectrum of the atmospheric eddies as shown below.

From the logarithmic wind profile relationship (Eq. 3) and the steady state fractional upward mass flux $f$ of surface air at any height $z$ (Eq. 4) the vertical velocity $W$ can be expressed as

$$W = w_* f z \tag{5}$$

The corresponding moisture content $q$ at height $z$ is related to the moisture content $q_*$ at the surface and is given as (from Eq. 5)

$$q = q_* f z \tag{6}$$

The aerosols are held in suspension by the eddy vertical velocity perturbations. Thus the suspended aerosol mass concentration $m$ at any level $z$ will be directly related to the vertical velocity perturbation $W$ at $z$, i.e., $W \sim mg$ where $g$ is the acceleration due to gravity. Therefore

$$m = m_* fz \quad (7)$$

In Eq. (7) $m_*$ is the suspended aerosol mass concentration in the surface layer. Let $r_a$ and $N$ represent the mean volume radius and number concentration of aerosols at level $z$. The variables $r_{as}$ and $N_*$ relate to corresponding parameters at the surface levels. Substituting for the average mass concentration in terms of mean radius and number concentration

$$\frac{4}{3}\pi r_a^3 N = \frac{4}{3}\pi r_{as}^3 N_* fz \quad (8)$$

The number concentration of aerosol decreases with height according to the $f$ distribution as shown earlier and is expressed as follows:

$$N = N_* f \quad (9)$$

## 3.2 Vertical variation of aerosol mean volume radius

The mean volume radius of aerosol increases with height as shown in the following.

The velocity perturbation $W$ is represented by an eddy continuum of corresponding size (length) scales $z$. The aerosol mass flux across unit cross-section per unit time is obtained by normalizing the velocity perturbation $W$ with respect to the corresponding length scale $z$ to give the volume flux of air equal to $Wz$ and can be expressed as follows from Eq. (5):

$$Wz = (w_* fz)z = w_* fz^2 \quad (10)$$

The corresponding normalised moisture flux perturbation is equal to $qz$ where $q$ is the moisture content per unit volume at level $z$. Substituting for $q$ from Eq. (6)

$$\text{normalised moisture flux at level } z = q_* fz^2 \quad (11)$$

The moisture flux increases with height resulting in increase of mean volume radius of cloud condensation nuclei CCN because of condensation of water vapour. The corresponding CCN (aerosol) mean volume radius $r_a$ at height $z$ is given in terms of the aerosol number concentration $N$ at level $z$ and mean volume radius $r_{as}$ at the surface as follows from Eq. (11)

$$\frac{4}{3}\pi r_a^3 N = \frac{4}{3}\pi r_{as}^3 N_* fz^2 \quad (12)$$

Substituting for $N$ from Eq. (9) in terms of $N_*$ and $f$

$$\begin{aligned} r_a^3 &= r_{as}^3 z^2 \\ r_a &= r_{as} z^{2/3} \end{aligned} \quad (13)$$

The mean aerosol size increases with height according to the cube root of $z^2$ (Eq. 13). As the large eddy grows in the vertical, the aerosol size spectrum extends towards larger sizes while the total number concentration decreases with height according to the $f$ distribution. The atmospheric aerosol size spectrum is dependent on the eddy energy spectrum and may be expressed in terms of the recently identified universal characteristics of fractal fluctuations generic to atmospheric flows as shown in Sec. 3.3 below.

## 3.3 Probability distribution of fractal fluctuations in atmospheric flows

The atmospheric eddies hold in suspension the aerosols and thus the size spectrum of the atmospheric aerosols is dependent on the vertical velocity spectrum of the atmospheric eddies. Atmospheric air flow is turbulent, i.e., consists of irregular fluctuations of all space-time scales characterized by a broadband spectrum of eddies. The suspended aerosols will also exhibit a broadband size spectrum closely related to the atmospheric eddy energy spectrum.

Atmospheric flows exhibit self-similar fractal fluctuations generic to dynamical systems in nature such as fluid flows, heart beat patterns, population dynamics, spread of forest fires, etc. Power spectra of fractal fluctuations exhibit inverse power law of form $f^{-\alpha}$ where $\alpha$ is a constant indicating long-range space-time correlations or persistence. Inverse power law for power spectrum indicates scale invariance, i.e., the eddy energies at two different scales (space-time) are related to each other by a scale factor ($\alpha$ in this case) alone independent of the intrinsic properties such as physical, chemical, electrical etc of the dynamical system.

Model predicted inverse power law form for atmospheric eddy energy spectrum is in agreement with earlier observational results (Selvam et al., 1992; Selvam and Joshi, 1995; Selvam et al., 1996; Selvam and Fadnavis, 1998; Selvam, 2011).

The general systems theory for turbulent fluid flows predicts that the eddy energy spectrum, i.e., the variance (square of eddy amplitude) spectrum is the same as the probability distribution $P$ of the eddy amplitudes, i.e. the vertical velocity $W$ values. Such a result that the additive amplitudes of eddies, when squared, represent the probabilities is exhibited by the subatomic dynamics of quantum systems such as the electron or photon. Therefore the unpredictable or irregular fractal space-time fluctuations generic to dynamical systems in nature, such as atmospheric flows is a signature of quantum-like chaos. The general systems theory for turbulent fluid flows predicts that the atmospheric eddy energy spectrum follows inverse power law form incorporating the *golden mean* $\tau$ and the normalized deviation $\sigma$ for values of $\sigma \geq 1$ and $\sigma \leq -1$ as given below

$$P = \tau^{-4\sigma} \qquad (14)$$

The vertical velocity $W$ spectrum will therefore be represented by the probability distribution $P$ for values of $\sigma \geq 1$ and $\sigma \leq -1$ given in Eq. (14) since fractal fluctuations exhibit quantum-like chaos as explained above.

$$W = P = \tau^{-4\sigma} \qquad (15)$$

Values of the normalized deviation $\sigma$ in the range $-1 < \sigma < 1$ refer to regions of primary eddy growth where the fractional volume dilution $k$ (Eq. 2) by eddy mixing process has to be taken into account for determining the probability distribution $P$ of fractal fluctuations (see Sec. 3.4 below).

## 3.4 Primary eddy growth region fractal space-time fluctuation probability distribution

Normalised deviation σ ranging from -1 to +1 corresponds to the primary eddy growth region. In this region the probability $P$ is shown to be equal to $P = \tau^{-4k}$ (see below) where $k$ is the fractional volume dilution by eddy mixing (Eq. 2).

The normalized deviation σ represents the length step growth number for growth stages more than one. The first stage of eddy growth is the primary eddy growth starting from unit length scale perturbation, the complete eddy forming at the tenth length scale growth, i.e., $R$ = 10$r$ and scale ratio $z$ equals 10. The steady state fractional volume dilution $k$ of the growing primary eddy by internal smaller scale eddy mixing is given by Eq. (2) as

$$k = \frac{w_* r}{WR} \tag{16}$$

The expression for $k$ in terms of the length scale ratio $z$ equal to $R/r$ is obtained from Eq. (1) as

$$k = \sqrt{\frac{\pi}{2z}} \tag{17}$$

A fully formed large eddy length $R$ = 10$r$ ($z$=10) represents the average or mean level zero and corresponds to a maximum of 50% probability of occurrence of either positive or negative fluctuation peak at normalized deviation σ value equal to zero by convention. For intermediate eddy growth stages, i.e., $z$ less than 10, the probability of occurrence of the primary eddy fluctuation does not follow conventional statistics, but is computed as follows taking into consideration the fractional volume dilution of the primary eddy by internal turbulent eddy fluctuations. Starting from unit length scale fluctuation, the large eddy formation is completed after 10 unit length step growths, i.e., a total of 11 length steps including the initial unit perturbation. At the second step ($z$ = 2) of eddy growth the value of normalized deviation σ is equal to 1.1 - 0.2 (= 0.9) since the complete primary eddy length plus the first length step is equal to 1.1. The probability of occurrence of the primary eddy perturbation at this σ value however, is determined by the fractional volume dilution $k$ which quantifies the departure of the primary eddy from its undiluted average condition and therefore represents the normalized deviation σ. Therefore the probability density $P$ of fractal fluctuations of the primary eddy is given using the computed value of $k$ as shown in the following equation.

$$P = \tau^{-4k} \tag{18}$$

The vertical velocity $W$ spectrum will therefore be represented by the probability density distribution $P$ for values of $-1 \le \sigma \le 1$ given in Eq. (18) since fractal fluctuations exhibit quantum-like chaos as explained above (Eq. 15).

$$W = P = \tau^{-4k} \tag{19}$$

The probabilities of occurrence (*P*) of the primary eddy for a complete eddy cycle either in the positive or negative direction starting from the peak value (σ = 0) are given for progressive growth stages (σ values) in the following Table 1. The statistical normal probability density distribution corresponding to the normalized deviation σ values are also given in the Table 1.

Table 1: Primary eddy growth

| Growth step no | ± σ | *k* | Probability (%) | |
|---|---|---|---|---|
| | | | Model predicted | Statistical normal |
| 2 | .9000 | .8864 | 18.1555 | 18.4060 |
| 3 | .8000 | .7237 | 24.8304 | 21.1855 |
| 4 | .7000 | .6268 | 29.9254 | 24.1964 |
| 5 | .6000 | .5606 | 33.9904 | 27.4253 |
| 6 | .5000 | .5118 | 37.3412 | 30.8538 |
| 7 | .4000 | .4738 | 40.1720 | 34.4578 |
| 8 | .3000 | .4432 | 42.6093 | 38.2089 |
| 9 | .2000 | .4179 | 44.7397 | 42.0740 |
| 10 | .1000 | .3964 | 46.6250 | 46.0172 |
| 11 | 0 | .3780 | 48.3104 | 50.0000 |

The model predicted probability density distribution *P* along with the corresponding statistical normal distribution with probability values plotted on linear and logarithmic scales respectively on the left and right hand sides are shown in Fig. 1. The model predicted probability distribution *P* for fractal space-time fluctuations is very close to the statistical normal distribution for normalized deviation σ values less than 2 as seen on the left hand side of Fig. 1. The model predicts progressively higher values of probability *P* for values of σ greater than 2 as seen on a logarithmic plot on the right hand side of Fig. 1.

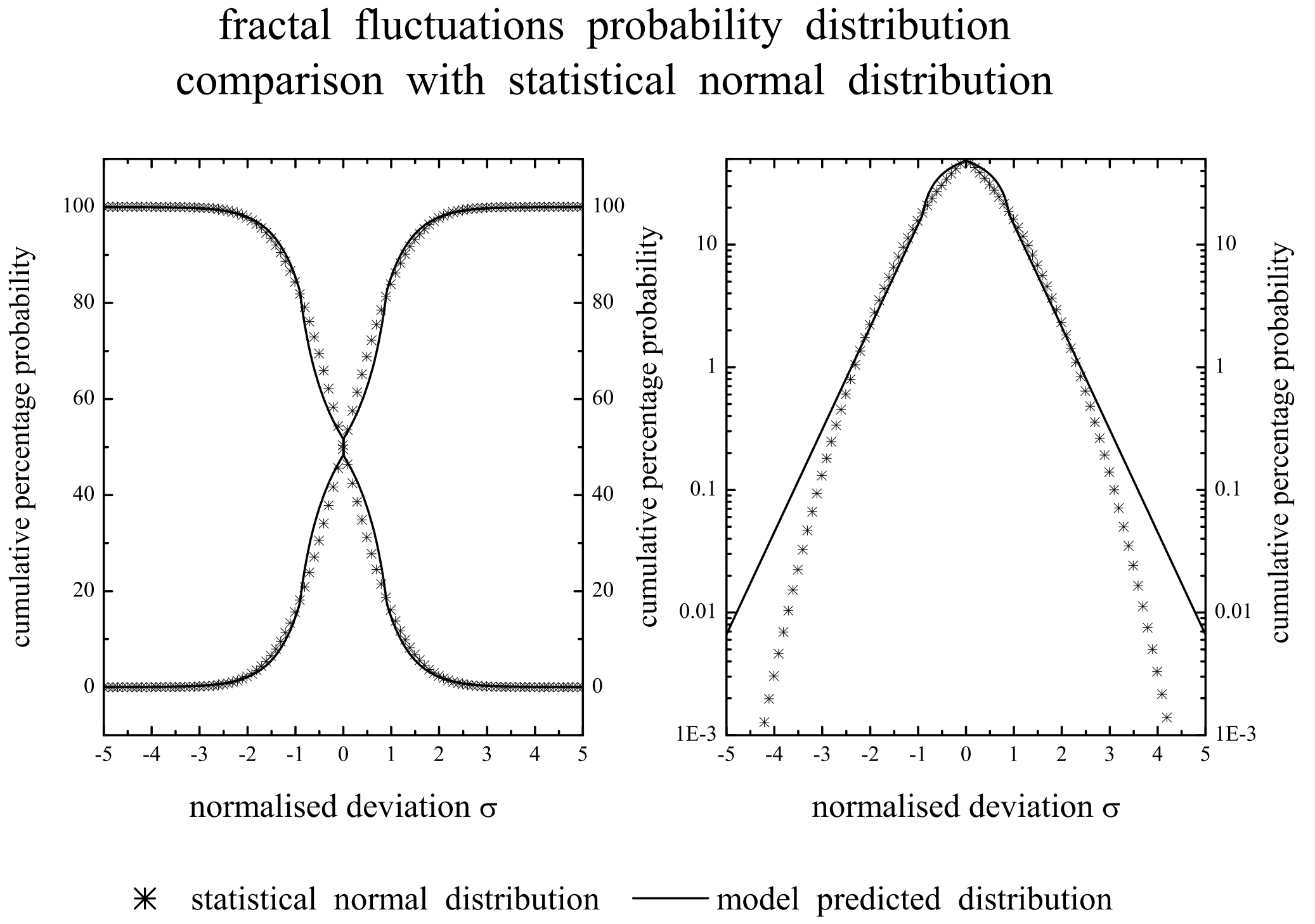

Fig. 1: Model predicted probability distribution *P* along with the corresponding statistical normal distribution with probability values plotted on linear and logarithmic scales respectively on the left and right hand sides.

## 3.5 Atmospheric wind spectrum and aerosol size spectrum

The steady state flux d$N$ of CCN at level $z$ in the normalised vertical velocity perturbation (d$W$)$z$ is given as

$$\mathrm{d}N = N(\mathrm{d}W)z \tag{20}$$

The logarithmic wind profile relationship for $W$ at Eq. (3) gives

$$\mathrm{d}N = Nz\frac{w_*}{k}\mathrm{d}(\ln z) \tag{21}$$

Substituting for $k$ from Eq. (2)

$$\mathrm{d}N = Nz\frac{w_*}{w_*}Wz\mathrm{d}(\ln z) = NWz^2\mathrm{d}(\ln z) \tag{22}$$

The length scale $z$ is related to the aerosol radius $r_a$ (Eq. 13). Therefore

$$\ln z = \frac{3}{2}\ln\left(\frac{r_a}{r_{as}}\right) \tag{23}$$

Defining a normalized radius $r_{an}$ equal to $\frac{r_a}{r_{as}}$, i.e., $r_{an}$ represents the CCN mean volume radius $r_a$ in terms of the CCN mean volume radius $r_{as}$ at the surface (or reference level). Therefore

$$\ln z = \frac{3}{2}\ln r_{an} \tag{24}$$

$$\mathrm{d}\ln z = \frac{3}{2}\mathrm{d}\ln r_{an} \tag{25}$$

Substituting for dlnz in Eq. (22)

$$\mathrm{d}N = NWz^2\frac{3}{2}\mathrm{d}(\ln r_{an}) \tag{26}$$

$$\frac{\mathrm{d}N}{\mathrm{d}(\ln r_{an})} = \frac{3}{2}NWz^2 \tag{27}$$

Substituting for $W$ from Eq. (15) and Eq. (19) in terms of the universal probability density $P$ for fractal fluctuations

$$\frac{\mathrm{d}N}{\mathrm{d}(\ln r_{an})} = \frac{3}{2}NPz^2 \tag{28}$$

The above equation is for the scale length $z$. The volume across unit cross-section associated with scale length $z$ is equal to $z$. The particle radius corresponding to this volume is equal to $z^{1/3}$

The equation (28) normalised for scale length and associated drop radius is given as

$$\frac{\mathrm{d}N}{\mathrm{d}(\ln r_{\mathrm{an}})} = \frac{3}{2}\frac{NPz^2}{z \times z^{\frac{1}{3}}} = \frac{3}{2}NPz^{\frac{2}{3}} \tag{28a}$$

The general systems theory predicts that fractal fluctuations may be resolved into an overall logarithmic spiral trajectory with the quasiperiodic Penrose tiling pattern for the internal structure such that the successive eddy lengths follow the Fibonacci mathematical number series. The eddy length scale ratio $z$ for length step $\sigma$ is therefore a function of the golden mean $\tau$ given as

$$z = \tau^{\sigma} \tag{29}$$

Expressing the scale length $z$ in terms of the golden mean $\tau$ in Eq. (28)

$$\frac{\mathrm{d}N}{\mathrm{d}(\ln r_{\mathrm{an}})} = \frac{3}{2}NP\tau^{\frac{2\sigma}{3}} \tag{30}$$

In Eq. (30) $N$ is the steady state aerosol concentration at level $z$. The normalized aerosol concentration any level $z$ is given as

$$\frac{1}{N}\frac{\mathrm{d}N}{\mathrm{d}(\ln r_{\mathrm{an}})} = \frac{3}{2}P\tau^{\frac{2}{3}\sigma} \tag{31}$$

The fractal fluctuations probability density is $P = \tau^{-4\sigma}$ (Eq. 15) for values of the normalized deviation $\sigma \geq 1$ and $\sigma \leq -1$ on either side of $\sigma = 0$ as explained earlier (Secs. 3.3, 3.4). Values of the normalized deviation $-1 \leq \sigma \leq 1$ refer to regions of primary eddy growth where the fractional volume dilution $k$ (Eq. 2) by eddy mixing process has to be taken into account for determining the probability density $P$ of fractal fluctuations. Therefore the probability density $P$ in the primary eddy growth region ($\sigma \geq 1$ and $\sigma \leq -1$) is given using the computed value of $k$ as $P = \tau^{-4k}$ (Eq. 19).

The normalised radius $r_{\mathrm{an}}$ is given in terms of $\sigma$ and the golden mean $\tau$ from Eq. (24) and Eq. (29) as follows.

$$\begin{aligned} \ln z &= \frac{3}{2}\ln r_{\mathrm{an}} \\ r_{\mathrm{an}} &= z^{2/3} = \tau^{2\sigma/3} \end{aligned} \tag{32}$$

The normalized aerosol size spectrum is obtained by plotting a graph of normalized aerosol concentration $\frac{1}{N}\frac{\mathrm{d}N}{\mathrm{d}(\ln r_{\mathrm{an}})} = \frac{3}{2}P\tau^{\frac{2\sigma}{3}}$ (Eq. 31) versus the normalized aerosol radius $r_{\mathrm{an}} = \tau^{2\sigma/3}$ (Eq. 32). The normalized aerosol size spectrum is derived directly from the universal probability density $P$ distribution characteristics of fractal fluctuations (Eq. 15 and Eq.19) and is independent of the height *z* of measurement and is universal for aerosols in turbulent atmospheric flows. The aerosol size spectrum is computed starting from the minimum size, the corresponding probability density *P* (Eq. 31) refers to the cumulative probability density starting from 1 and is computed as equal to $P = 1 - \tau^{-4\sigma}$.

The universal normalised aerosol size spectrum represented by $\frac{1}{N}\frac{\mathrm{d}N}{\mathrm{d}(\ln r_{an})}$ versus $r_{an}$ is shown in Fig. 2.

Model predicted aerosol size spectrum

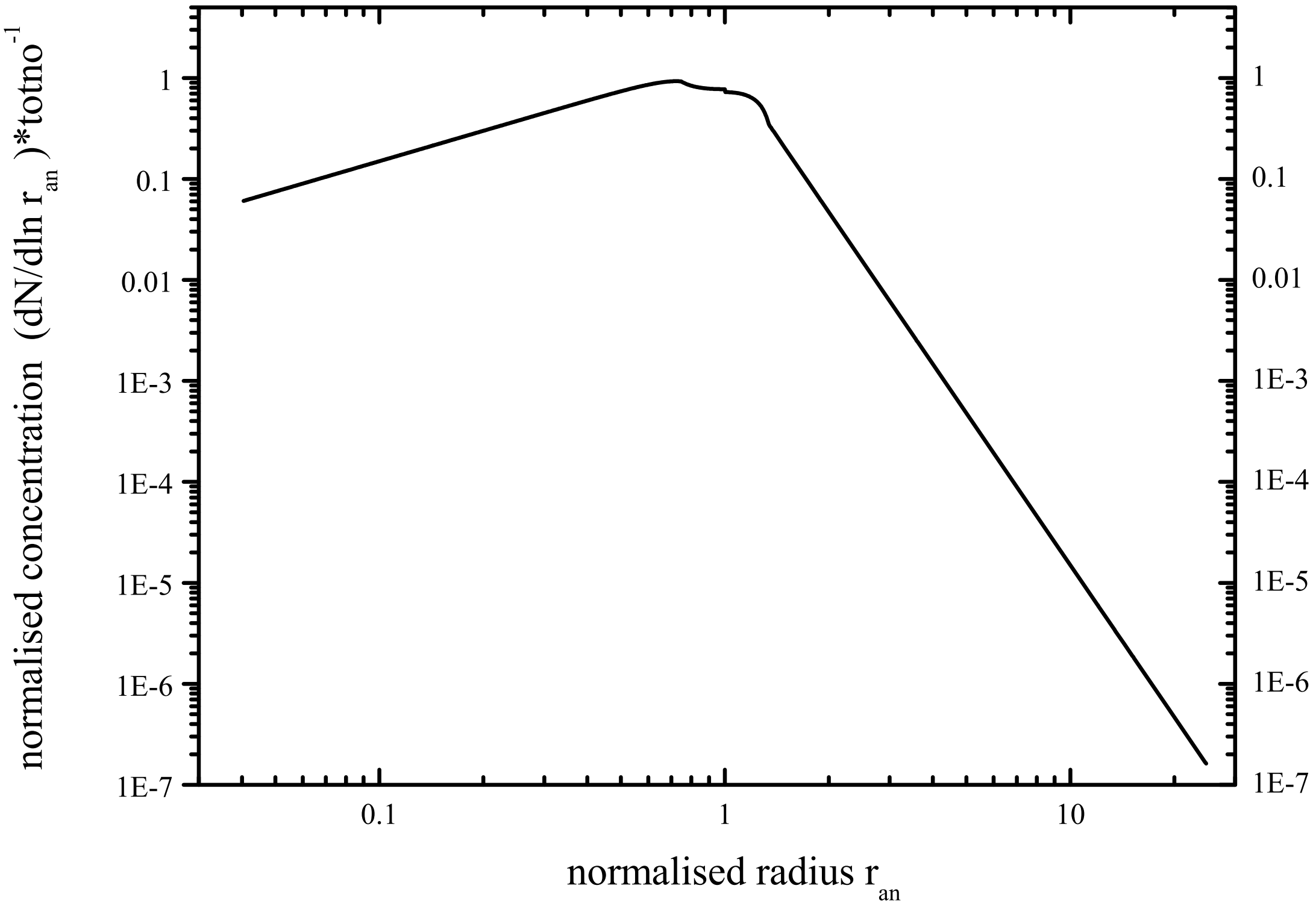

Fig. 2: Model predicted aerosol size spectrum

### 3.6 General systems theory and maximum entropy principle

The maximum entropy principle concept of classical statistical physics is applied to determine the fidelity of the inverse power law probability distribution *P* (Eq. 15 and Eq. 19) for exact quantification of the observed space-time fractal fluctuations of dynamical systems ranging from the microscopic dynamics of quantum systems to macro-scale real world systems. Kaniadakis (2009) states that the correctness of an analytic expression for a given power-law tailed distribution used to describe a statistical system is strongly related to the validity of the generating mechanism. In this sense the maximum entropy principle, the cornerstone of statistical physics, is a valid and powerful tool to explore new roots in searching for generalized statistical theories (Kaniadakis, 2009). The concept of entropy is fundamental in the foundation of statistical physics. It first appeared in thermodynamics through the second law of thermodynamics. In statistical mechanics, we are interested in the disorder in the distribution of the system over the permissible microstates. The measure of disorder first provided by Boltzmann principle (known as Boltzmann entropy) is given by *S* =

$K_B \ln M$, where $K_B$ is the thermodynamic unit of measurement of entropy and is known as Boltzmann constant. $K_B = 1.38 \times 10^{-16}$ erg/°C. $M$, called thermodynamic probability or statistical weight, is the total number of microscopic complexions compatible with the macroscopic state of the system and corresponds to the "degree of disorder" or 'missing information' (Chakrabarti and De, 2000). For a probability distribution among a discrete set of states the generalized entropy for a system out of equilibrium is given as (Salingaros and West, 1999; Chakrabarti and De, 2000; Beck, 2009; Sethna, 2009).

$$S = -\sum_{j=1}^{\sigma} P_j \ln P_j \qquad (33)$$

In Eq. (33) $P_j$ is the probability for the $j^{th}$ stage of eddy growth in the present study, σ is the length step growth which is equal to the normalized deviation and the entropy $S$ represents the 'missing information' regarding the probabilities. Maximum entropy $S$ signifies minimum preferred states associated with scale-free probabilities.

The validity of the probability distribution $P$ (Eq. 15 and Eq. 19) is now checked by applying the concept of maximum entropy principle (Kaniadakis, 2009). The probability distribution $P$ is equal to -log$P$ as shown in the following

The r.m.s circulation speed $W$ of the large eddy follows a logarithmic relationship with respect to the length scale ratio $z$ equal to $R/r$ (Eq. 3 ) as given below

$$W = \frac{w_*}{k} \log z$$

In the above equation the variable $k$ represents for each step of eddy growth, the fractional volume dilution of large eddy by turbulent eddy fluctuations carried on the large eddy envelope (Selvam, 1990, 2007) and is given as (Eq. 16)

$$k = \frac{w_* r}{WR}$$

Substituting for $k$ in Eq. (16) we have

$$W = w_* \frac{WR}{w_* r} \log z = \frac{WR}{r} \log z$$
$$and \qquad (34)$$
$$\frac{r}{R} = \log z$$

The ratio $r/R$ represents the fractional probability $P$ of occurrence of small-scale fluctuations ($r$) in the large eddy ($R$) environment. Since the scale ratio $z$ is equal to $R/r$, Eq. 34) may be written in terms of the probability $P$ as follows.

$$\frac{r}{R} = \log z = \log\left(\frac{R}{r}\right) = \log\left(\frac{1}{(r/R)}\right)$$
$$P = \log\left(\frac{1}{P}\right) = -\log P \tag{35}$$

Substituting for log $P_j$ (Eq. 35) and for the probability $P_j$ in terms of the golden mean $\tau$ derived earlier (Eq. 15 and Eq. 19) the entropy $S$ (Eq. 33) is expressed as

$$S = -\sum_{j=1}^{\sigma} P_j \log P_j = \sum_{j=1}^{\sigma} P_j^2 = \sum_{j=1}^{\sigma} \left(\tau^{-4\sigma}\right)^2$$
$$S = \sum_{j=1}^{\sigma} \tau^{-8\sigma} \approx 1 \text{ for large } \sigma \tag{36}$$

In Eq. (36) $S$ is equal to the square of the cumulative probability density distribution and it increases with increase in $\sigma$, i.e., the progressive growth of the eddy continuum and approaches 1 for large $\sigma$. According to the second law of thermodynamics, increase in entropy signifies approach of dynamic equilibrium conditions with scale-free characteristic of fractal fluctuations and hence the probability distribution $P$ (Eq. 15 and Eq. 19) is the correct analytic expression quantifying the eddy growth processes visualized in the general systems theory. The ordered growth of the atmospheric eddy continuum is associated with maximum entropy production.

Paltridge (2009) states that the principle of maximum entropy production (MEP) is the subject of considerable academic study, but is yet to become remarkable for its practical applications. The ability of a system to dissipate energy and to produce entropy ”ought to be″ some increasing function of the system’s structural complexity. It would be nice if there were some general rule to the effect that, in any given complex system, the steady state which produces entropy at the maximum rate would at the same time be the steady state of maximum order and minimum entropy (Paltridge, 2009).

Earlier studies on the application of the concept of maximum entropy in atmospheric physics are given below. A systems theory approach based on maximum entropy principle has been applied in cloud physics to obtain useful information on droplet size distributions without regard to the details of individual droplets (Liu et al. 1995; Liu 1995; Liu and Hallett 1997, 1998; Liu and Daum, 2001; Liu, Daum and Hallett, 2002; Liu, Daum, Chai and Liu, 2002). Liu, Daum et al. (2002) conclude that a combination of the systems idea with multiscale approaches seems to be a promising avenue. Checa and Tapiador (2011) have presented a maximum entropy approach to Rain Drop Size Distribution (RDSD) modelling. Liu, Liu and Wang (2011) have given a review of the concept of entropy and its relevant principles, on the organization of atmospheric systems and the principle of the Second Law of thermodynamics, as well as their applications to atmospheric sciences. The Maximum Entropy Production Principle (MEPP), at least as used in climate science, was first hypothesized by Paltridge (1978).

# 4. Data

Daily mean volume particle size distribution d$V$($r$)/dln$r$ ($\mu m^3/\mu m^2$) retrieved in 22 logarithmically equidistant bins in the range of sizes 0.05μm $\leq$ r $\leq$ 15 μm for Davos (Switzerland) and Mauna Loa (Hawaii) for the year 2010 and Izana (Spain) for the year 2009 were obtained from AERONET aerosol robotic network (http://aeronet.gsfc.nasa.gov/new_web/data.html). For Izana (Spain) the complete data set for the 12 month period of 2009 were available and hence it was used. Daily Average data are calculated from all points for each day when three (3) or more points are available. Data sets for a total of 54, 180 and 133 days respectively were available for the three stations Davos, Mauna Loa and Izana. The formulas for calculating standard parameters of the particle size distribution (http://aeronet.gsfc.nasa.gov/new_web/Documents/Inversion_products_V2.pdf) are given below.

AERONET retrieves the aerosol size distribution of the particle volume d$V(r)$/dln$r$. It relates to the distribution of particle number as follows:

$$\frac{\mathrm{d}V(r)}{\mathrm{d}\ln(r)} = V(r)\frac{\mathrm{d}N(r)}{\mathrm{d}\ln(r)} = \frac{4}{3}\pi r^3 \frac{\mathrm{d}N(r)}{\mathrm{d}\ln(r)}$$

Volume concentration (μm³/μm²):

$$C_v = \int_{r_{min}}^{r_{max}} \frac{\mathrm{d}V(r)}{\mathrm{d}\ln(r)}$$

The details of computations in the present study are as follows:

i. The mid-point (center of size bin) radius $r$ was used to calculate the distribution of particle number d$N(r)$/dln($r$) from the volume concentration d$V(r)$/dln($r$) for fine (f) and course (c) aerosol modes.

ii. The radius range for fine (f) mode is $0.05 < r \le 0.6$ μm and the radius range for the coarse (c) mode is $0.6 < r \le 15$ μm.

iii. The total number concentrations $N$ ($\sum \mathrm{d}N(r)$) for fine (f) and coarse (c) modes were calculated from d$N(r)$/dln($r$) since dln($r$) is a constant equal to d($r$)/$r$ for the retrieved size spectrum with logarithmically equidistant bins. The constant dln($r$) computed as equal to $(r_2-r_1)/r_1$ from the known values of mid-point (center of size bin) radii $r$ is equal to 0.31207.

# 5. Analysis and Results

The atmospheric suspended particulate size spectrum is closely related to the vertical velocity spectrum (Sec. 3). The mean volume radius of suspended aerosol particulates increases with height (or reference level $z$) in association with decrease in number concentration. At any height (or reference level) $z$, the fractal fluctuations (of wind, temperature, etc.) carry the signatures of eddy fluctuations of all size scales since the eddy of length scale $z$ encloses smaller scale eddies and at the same time forms part of internal circulations of eddies larger than length scale $z$ (Sec. 3.2). The observed atmospheric suspended particulate size spectrum also exhibits a decrease in number concentration with increase in particulate radius. At any reference level $z$ of measurement the mean volume radius $r_{as}$ will serve to calculate the normalized radius $r_{an}$ for the different radius class intervals as explained below.

The general systems theory for fractal space-time fluctuations in dynamical systems predicts universal mass size spectrum for atmospheric suspended particulates (Sec. 3). For homogeneous atmospheric suspended particulates, i.e. with the same particulate substance density, the atmospheric suspended particulate mass and radius size spectrum is the same and is given as (Sec. 3.5) the normalized aerosol number concentration equal to $\frac{1}{N}\frac{dN}{d(\ln r_{an})}$ versus the normalized aerosol radius $r_{an}$, where (i) $r_{an}$ is equal to $\frac{r_a}{r_{as}}$, $r_a$ being the mean class interval radius and $r_{as}$ the mean volume radius for the total aerosol size spectrum (ii) $N$ is the total aerosol number concentration and d$N$ is the aerosol number concentration in the aerosol radius class interval d$r_a$ (iii) d(ln $r_{an}$) is equal to $\frac{d r_a}{r_a}$ for the aerosol radius class interval $r_a$ to $r_a$+d$r_a$.

The average normalized aerosol size spectra for fine (f) mode for Davos, Mauna Loa and Izana with 54, 180 and 133 daily mean data sets respectively are shown in Fig. 3 along with the model predicted universal normalized aerosol size spectrum. The corresponding aerosol size spectra for coarse (c) mode are given in Fig. 4. The observed aerosol size spectra are in close agreement with model predicted universal spectrum for suspended particulates in the turbulent atmospheric flows. The total average mean volume radius and total number concentration for the three stations for the period of study are given in Fig. 5. The mean volume radius and total number concentration are minimum for Mauna Loa (Hawaii) for both fine and coarse aerosol modes. Coarse mode particulate number concentration is a maximum for Izana (Spain).

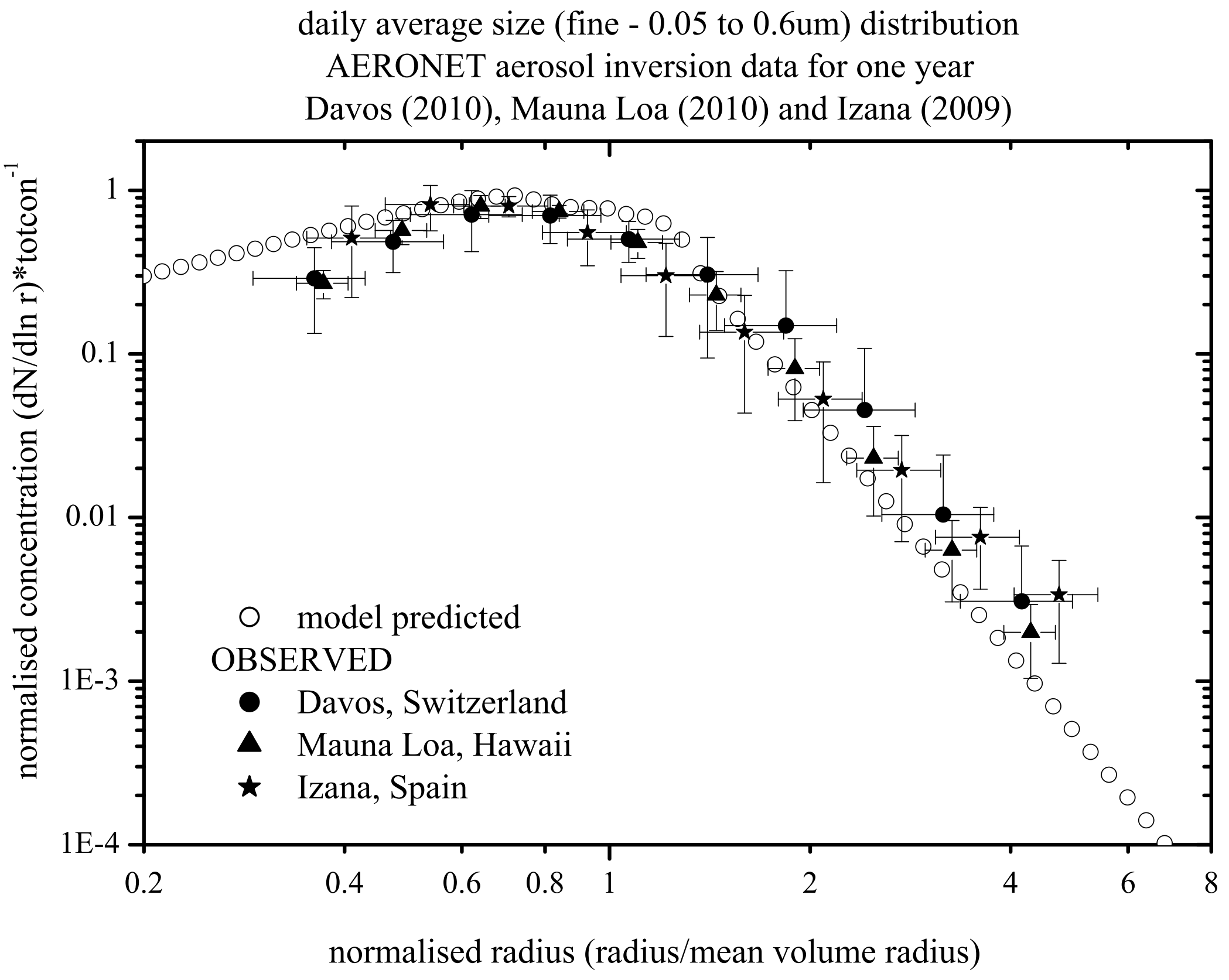

Fig. 3. Mean daily average aerosol size distribution, fine mode (0.05 < radius ≤ 0.6 μm) for Davos and Mauna Loa for the year 2010 and Izana for the year 2009.

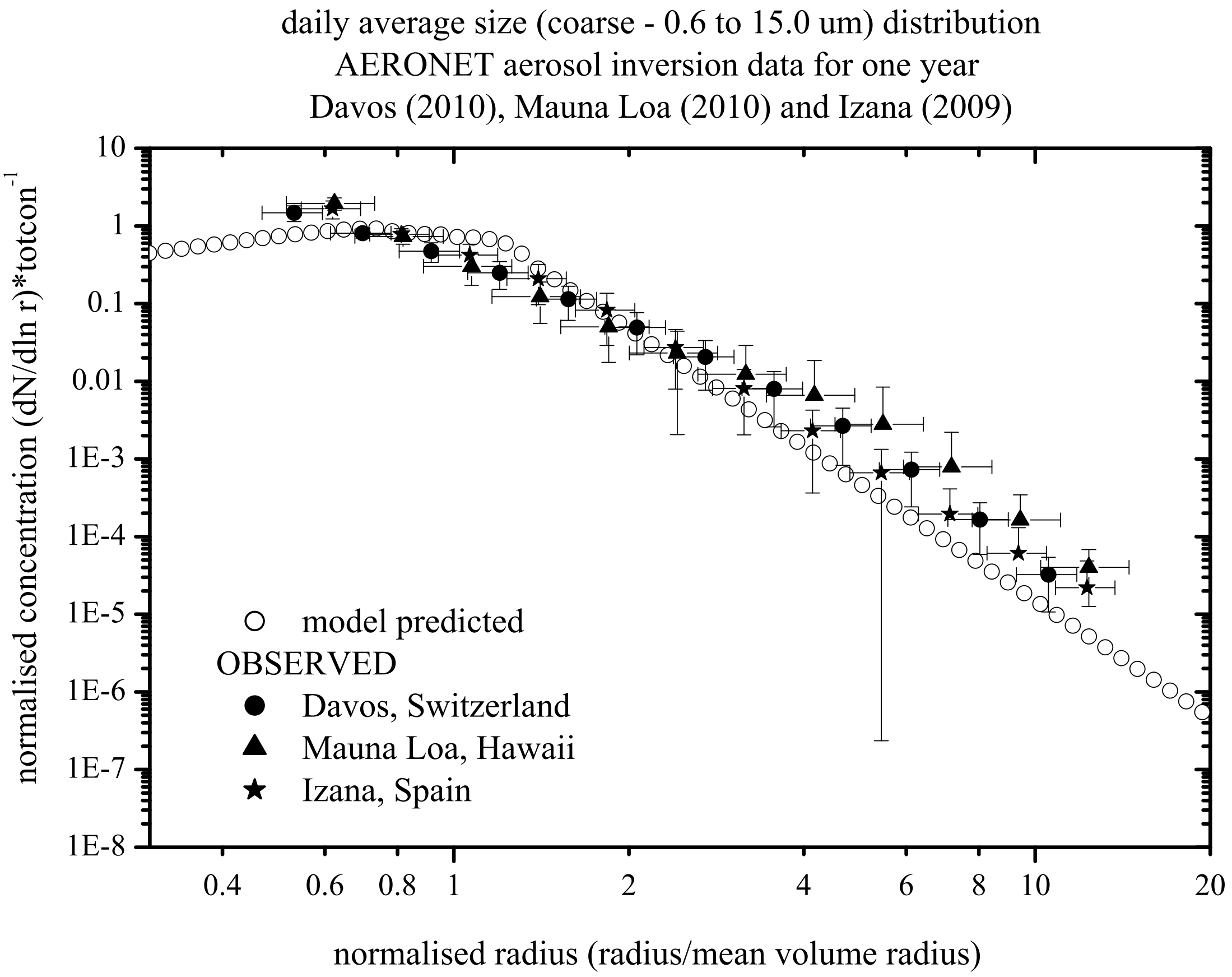

Fig. 4. Mean daily average aerosol size distribution, coarse mode (0.6 < radius ≤ 15.0 μm) for Davos and Mauna Loa for the year 2010 and Izana for the year 2009.

AERONET aerosol inversion data (daily averages) for one year
average mean volume radius and total number concentration
for fine (0.05 - 0.6 um) and coarse (0.6 to 15.0 um) aerosol
Davos (2010), Mauna Loa (2010) and Izana (2009)

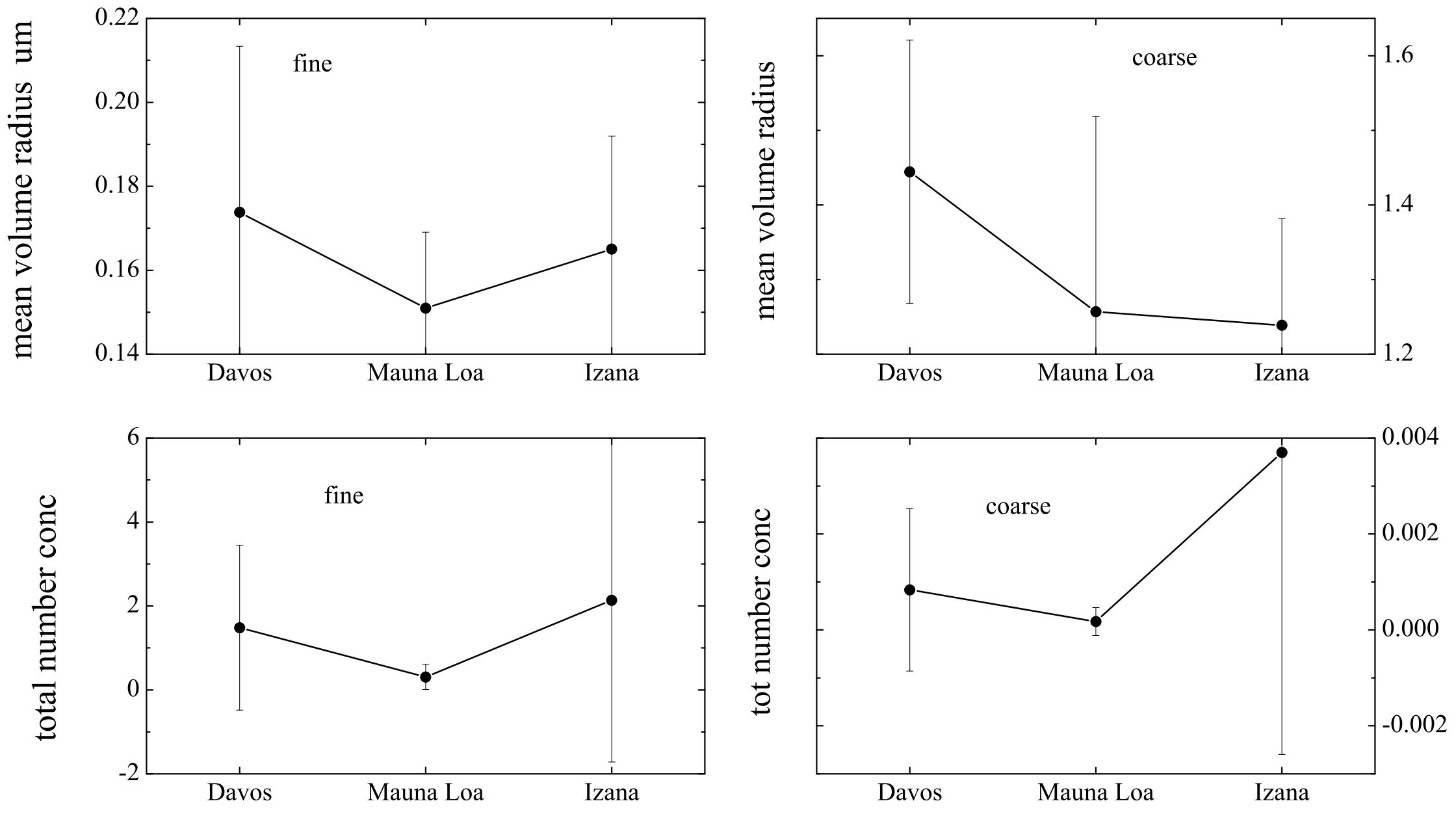

Fig. 5. The total average mean volume radius and total number concentration for Davos and Mauna Loa for the year 2010 and Izana for the year 2009, the corresponding number of daily average data sets being 54, 180 and 133.

# 6. Discussion and Conclusions

Atmospheric flows exhibit selfsimilar fractal fluctuations on all space-time scales. Fractal fluctuations are ubiquitous to dynamical systems in nature such as fluid flows, heart beat patterns, population growth etc. Power spectra of fractal fluctuations exhibit inverse powerlaw form indicating long-range correlations, identified as self-organised criticality. Identification and quantification of the exact physical laws underlying the observed selforganised criticality will help predict the future evolution of dynamical systems such as weather patterns. A general systems theory which satisfies the maximum entropy principle of classical statistical physics recently proposed by the author enables formulation of precise quantitative relations for the observed universal characteristics of fractal fluctuations in turbulent atmospheric flows. The model predictions are as follows. (i) The apparently chaotic (unpredictable) fluctuations can be resolved into a nested continuum of vortex roll circulations tracing the space filling quasiperiodic Penrose tiling pattern with an overall logarithmic spiral trajectory. (ii) The amplitude and also the variance (square of the amplitude) of fractal fluctuations are quantified by the same statistical probability distribution function incorporating the golden mean $\tau$ and exhibits scale-free universal inverse powerlaw characteristics. Therefore, fractal fluctuations are signatures of quantumlike chaos since square of the eddy amplitudes, i.e., variances, represent the probability densities (of amplitudes), a property exhibited by the subatomic dynamics of quantum systems such as electron or photon. (iii) Atmospheric particulates are held in suspension in the fractal fluctuations of vertical wind velocity. The mass or radius (size) distribution for homogeneous suspended atmospheric particulates is expressed as a universal scale-independent function of the golden mean τ, the total number concentration and the mean volume radius. The universal aerosol size spectrum will have applications in computations of radiation balance of earth-atmosphere system in climate models.

Model predictions are in agreement separately, with the fine (0.1 to 0.6 μm) and the coarse (0.6 to 15 μm) mode AERONET aerosol inversion data sets (daily averages) for Davos and Mauna Loa for the year 2010 and Izana for 2009. The results indicate two different homogeneous aerosol substance densities corresponding to the two size (radius) ranges, namely (i) 0.1 to 0.6 μm (fine mode) and (ii) 0.6 to 15 μm (coarse mode). Earlier studies (Hussar, 2005) have shown that the source for submicron (diameter) size accumulation mode aerosols is different from the larger (greater than 1 μm diameter) coarse mode particles in the atmosphere and therefore may form two different homogeneous aerosol size groups.

The source of the uncertainties displayed by the error bars (Figs. 3 and 4 for total average) may be due to measurement noise, independent in every size interval, also may be due to different aerosol sources. The model predicts universal spectrum for suspended aerosol mass size distribution (Sec. 3), based on the concept that the atmospheric eddies hold in suspension the aerosols and thus the mass size spectrum of the atmospheric aerosols is dependent on the vertical velocity fluctuation spectrum of the atmospheric eddies.

At present empirical models such as the log normal distribution with arbitrary constants for the size distribution of atmospheric suspended particulates are used for quantitative estimation of earth-atmosphere radiation budget related to climate warming/cooling trends. The general systems theory model for aerosol size distribution is scale free and is derived directly from atmospheric eddy dynamical concepts. The universal aerosol size spectrum, a function of the total aerosol number concentration and the mean volume radius alone, will have applications in computations of radiation balance of earth-atmosphere system in climate models.

## Acknowledgements

The author is grateful to Dr. Christoph_Wehrli, Dr. Brent. N. Holben and Dr. Philippe Goloub, the respective Principal Investigators of AERONET sites Davos, Mauna Loa and Izana and their team members for free and generous distribution of aerosol data to the scientific community and their efforts in establishing and maintaining the AERONET sites. The author is grateful to Dr. A. S. R. Murty for encouragement during the course of the study.